\documentclass[%
 twocolumn,
 amsmath,amssymb,
 aps,
]{revtex4-1}

\usepackage{graphicx}
\usepackage{dcolumn}
\usepackage{bm}
\usepackage[parfill]{parskip}
\usepackage{epstopdf}
\usepackage{float}
\usepackage{tikz}
\usetikzlibrary{arrows}
\usepackage{xcolor}
\usepackage{braket}

\definecolor{dred}{HTML}{AE2D2D}

\begin{document}

\title{Design, fabrication and characterisation of a micro-fabricated double-junction segmented ion trap}
\author{Chiara Decaroli}%
\author{Roland Matt}
\author{Robin Oswald}
\author{Maryse Ernzer}
\author{Jeremy Flannery}
\author{Simon Ragg}
\author{Jonathan P. Home}%
\affiliation{%
Trapped Ion Quantum Information Group, Institute for Quantum Electronics,\\ ETH Zurich, 8093 Zurich,\\ Switzerland}


\date{\today}

\begin{abstract}
We describe the implementation of a three-dimensional Paul ion trap fabricated from a stack of precision-machined silica glass wafers, which incorporates a pair of junctions for 2-dimensional ion transport. The trap has 142 dedicated electrodes which can be used to define multiple potential wells in which strings of ions can be held. By supplying time-varying potentials, this also allows for transport and re-configuration of ion strings. We describe the design, simulation, fabrication and packaging of the trap, including explorations of different parameter regimes and possible optimizations and design choices. We give results of initial testing of the trap, including measurements of heating rates and junction transport. 

\end{abstract}

\maketitle

\section{INTRODUCTION}\label{Sec:Intro}
Trapped ions are among the leading systems for realizing  quantum information processing \cite{Leibfried2003}, with prior demonstrations including high quality operations \cite{Hughes2020, fastOxford2018, vlad2018} as well as incorporation of multiple features required for scaling \cite{Karan2020, Niffenegger2020, Todaro2021}. However, useful and practical quantum computers are expected to require ion numbers on the order of 10$^6$ \cite{Kielpinski2002architecture, Monroe_optilink2013, natureCirac2000}, which is far beyond current experimental capabilities. Thus an important area of research is how to interconnect elementary quantum processors to build a larger scale system.  There are two main approaches towards this scaling challenge: one is to connect individual traps via optical links \cite{Monroe_optilink2013, Stephenson2020_optilink}, while the second uses the transport of ions within a single device to connect a large array \cite{Kielpinski2002architecture}. 
The quantum-CCD approach is demanding due to the need to implement traps with complex segmented electrode structures, the need to connect these to dynamic control, and their integration with optical delivery. In addition, a central challenge in realising high connectivity between qubits is to implement some means of re-configuring ion chains. This can be performed to a limited extend using rotations of ion chains \cite{shuttlingrot2020,rotationsion2020,ionswap2017}, but a more flexible approach is to implement 2-dimensional junctions, where an arm of the junction is chosen for ion transport according to the desired protocol, and implemented by the choice of dynamic potentials applied to the electrodes \cite{transportNIST2020,Tjunction2006, Moehring2011_surfacejunc, Wright2013_surfacejunc,shu2014_Ysurfacejunc}. To date, traps incorporating junctions have been successfully fabricated and operated using  3-dimensional electrode structures and with electrodes laid out on a 2-dimensional surface. These all realize intersections of linear trapping regions, which take either an X, T, or Y shape depending on the implementation. Transporting ions through junctions
is challenging because ions must transition through regions in which they experience significant driven micromotion from the radio-frequency drive of the trap; the pseudopotential in these regions is non-zero. Imperfections in the transport can therefore be transformed into instabilities or cause undesired excitation of ions. Experimental work using junctions has had mixed results. At NIST, using 3-dimensional electrode structures,  transport of ions into both arms of the junction was performed with observed heating of less than a quanta per transport \cite{Blakestad2009high, blakestad2011, transportNIST2020}, which was a significant improvement on earlier attempts \cite{Tjunction2006}. In that work, the performance was achieved despite the ions traversing a 0.3 electron-volt pseudopotential barrier during transport, but required additional filtering of the radio-frequency drive in order to avoid excess heating. In surface-electrode traps many attempts to produce and operate junctions have been made, with the flexibility of fabrication of small-scale electrode structures allowing geometries designed to minimize the pseudopotential barriers \cite{Moehring2011_surfacejunc, Wright2013_surfacejunc,shu2014_Ysurfacejunc,sandiaHOAtrap}. Nevertheless, the  best reported performance in terms of heating per transport is between 40 and 150 quanta \cite{Wright2013_surfacejunc}. While mature fabrication techniques mean that surface-electrode traps are easier to manufacture at high complexity \cite{sussexTrapsReview2020} and thus provide additional freedom in design, they are accompanied by challenges in control. This is primarily due to the naturally asymmetric geometry around the ion position which reduces trap depth and curvature and produces anharmonicity \cite{Home2011Anharmonic}. With free-space laser delivery they are also susceptible to stray electric fields caused by laser light scattering off the electrodes \cite{surfaceTrapCharging2011}. 

In this article, we present a trap with two junctions which takes advantage of recent developments in fused silica micro-fabrication \cite{simon2019} to realize the 3-dimensional structures required. This fabrication technique introduces additional flexibility compared to previous laser ablation of alumina and aluminium nitride wafers. We examine how this can be utilized to optimize the geometry of the electrodes, in particular with regards to the junction, exploring the trade-off between pseudopotential barrier height and gradient and confinement at the center. Using the ability to precisely align the wafers \cite{simon2019}, we pursue a trap with trapping regions throughout its 9~mm length. We describe the design, simulation and packaging of the trap, as well as initial characterization of trap performance using trapped calcium ions. \\
The article is structured as follows: section \ref{sec:trapdesign} gives an overview of the trap design given the constraints of the fabrication process. Section \ref{sec:simulations} describes the simulations of the trap, Section \ref{sec:fabrication} details the fabrication process. Section \ref{sec:measurements} presents the experimental apparatus and the experimental verification of the trap, including heating rate measurements, characterization of axial micromotion, and initial transport tests.

\section{Trap design}\label{sec:trapdesign}
The trap takes advantage of precision laser machining using laser-enhanced etching of SiO$_2$. The manufacturing technique and its use for producing precisely machined traps using stacked wafers has been described previously \cite{simon2019}. Here we review the considerations relevant to the particular trap described, as well as give an overview of the full electrode structure. 

\subsection{Trap overview}\label{Sec:TrapOverview}

\begin{figure}[h]
\centering
\includegraphics[scale=0.65]{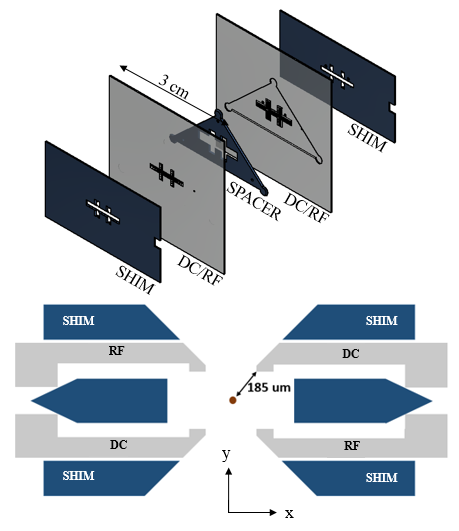}
\caption{Top: the 5 wafers that compose the trap stack.
The main trapping wafers (shown in grey) are 30$\times$30 mm in size, and 300 $\mu$m thick. Potentials applied to compensation electrodes situated above and below the main trapping wafers (shown in blue) are used to cancel stray electric fields, while the central wafer (spacer) is used for alignment. Bottom: side view of the stack. The ion-electrode distance is 185 $\mu$m. }
\label{fig:trap stack}
\end{figure}

\begin{figure*}
\centering
\includegraphics[scale=0.6]{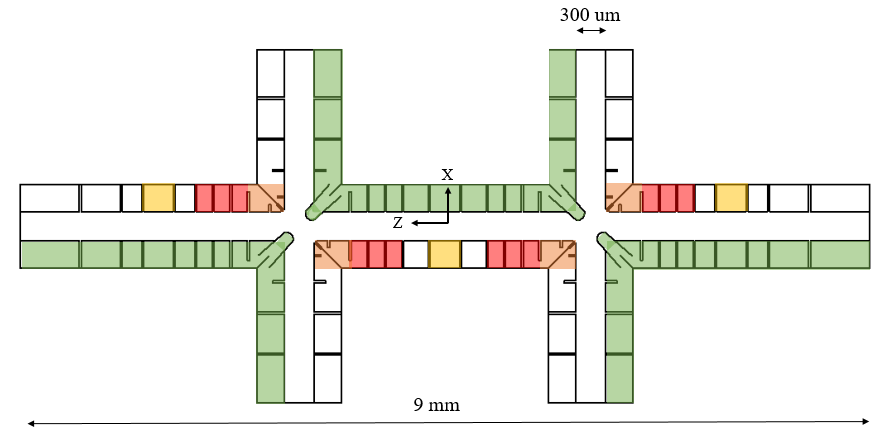}
\caption{Top view of one of the trapping wafers. The RF electrode is shaded in green. The experimental zones are adjacent to DC electrodes shaded in yellow,  and the splitting and recombination zones are adjacent to the electrodes shaded red. The DC electrodes used for control at the junction centers are shaded in orange.}
\label{fig:trapZones}
\end{figure*}

Femtosecond laser machining allows on one hand for the fabrication of precise features which can be used as alignment guides, on the other for the creation of self-standing arbitrary 3-dimensional structures \cite{simon2019}. We focus on harnessing this fabrication technique to produce a 3-dimensional micro-machined ion trap which includes three individual laser control regions connected via two X junctions whose legs may be used as storage areas, and two loading zones. \\
The trap is made of a stack of five machined silica glass wafers, as shown in Figure~\ref{fig:trap stack}. It incorporates a number of precisely machined features. These have been previously discussed in detail in \cite{simon2019} and include micron-level precise alignment guides for trap assembly, tapered electrode fingers for optical access, 3-dimensional protruding electrodes at the junctions and vias to route electrodes from the top to the bottom surface of each wafer. The inner wafer serves as a spacer and alignment piece. It is sandwiched by the two main electrode wafers hosting RF and DC electrodes. The outer layers of the sandwich host further ``shim'' DC electrodes, used for stray field compensation. The electrode structure is sketched in Figure \ref{fig:trapZones}. It is about 9 mm long and can be separated into three axial sections, the outer two each contain a loading zone, an ``experiment'' zone dedicated for laser control of ions, and a splitting region. The central region (between the junctions) consists of two splitting regions and one experiment zone. The electrodes are separated in the $y$ direction by 220~$\mu$m. In the plane of the wafers ($x-z$) the separations of RF and DC electrodes are 300 $\mu$m, resulting in a 185 $\mu$m electrode-ion distance and a 82 degree opening angle. At the junction, two smaller DC electrodes (140 $\mu$m wide) provide additional degrees of freedom. There are 106 DC control electrodes and 36 DC shim electrodes, for a total of 142 DC signal lines and one RF electrode.

\section{Trap simulation}\label{sec:simulations}
We simulated the trap as well as smaller subsections. This included studies of the ability to mitigate residual axial fields originating from the junction structures and the finite size of electrodes through control of electrode shapes, as well as wafer misalignment. 

\begin{figure}[h]
\centering
\includegraphics[scale=0.4]{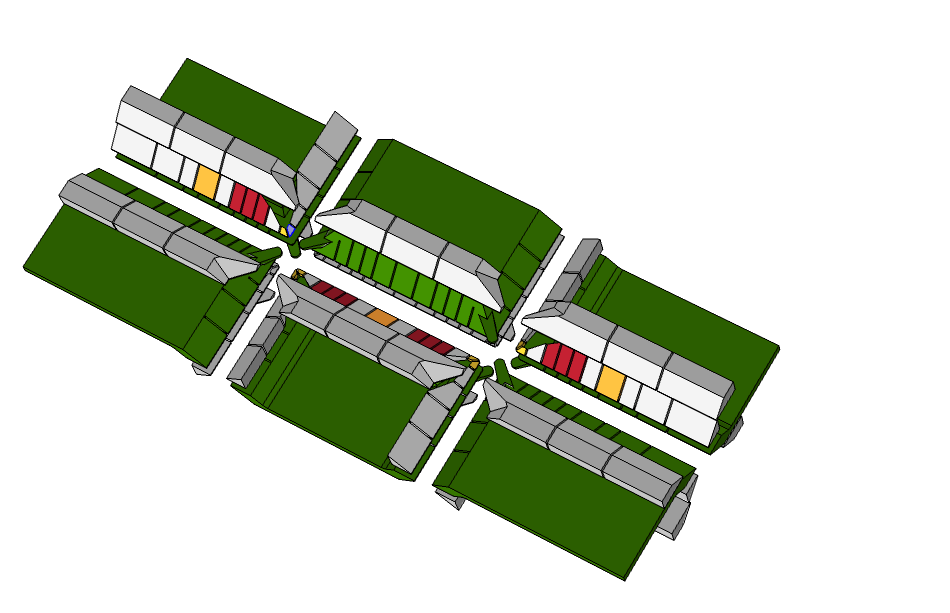}
\caption{Geometric model of the trap used to compute the electric fields and electric potential in COMSOL. To guide the reader, the RF electrode is here shown in green, the experimental zone DC electrodes in yellow and the splitting DC electrodes in red, as in Figure 2.}
\label{fig:trapModel}
\end{figure}

\begin{figure}[h]
\centering
\includegraphics[scale=0.4]{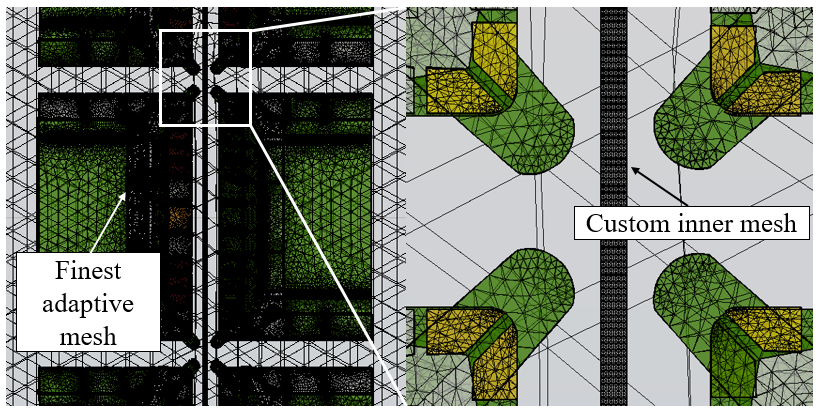}
\caption{Detail of the meshing geometry. Free tetrahedrons are used to mesh the entire trap volume. A smaller rectangular airbox is placed close to the region of interest, on which we impose a custom mesh with maximum element size of 2 $\mu$m.}
\label{fig:mesh}
\end{figure}

The trap is simulated using the finite element method
(FEM) software COMSOL Multiphysics. The simulation is performed for each electrode separately. Each round consists of setting the chosen electrode to 1 V with all others grounded, and calculating the electric field components over a region of interest. The data is then extracted over a chosen grid. The trap model used for the simulation is shown in Figure \ref{fig:trapModel}. The electrodes are enclosed in a grounded airbox (10$\times$5$\times$15 mm, not shown), which serves to define the boundary conditions of the electrostatic problem. The geometry is meshed using the highest resolution in-built COMSOL meshing algorithm within the airbox (adaptive mesh with maximum element size = 0.3 mm, minimum element size = 30 $\mu$m). An additional airbox enclosing a small area of interest near the trapping location is defined and a custom mesh is applied within this box. Here the maximum size of a mesh element is chosen to be 1 or 2 $\mu$m. Smaller mesh sizes, while drastically increasing computational resources, do not provide better resolution for our geometry. Figure \ref{fig:mesh} illustrates the results of the FEM meshing algorithm close to one of the junctions, and shows the smaller custom airbox defined around the trapping area. Another possible approach is to use a Boundary Element Method solver (BEM) \cite{NumericalTools2010}. In this case, the optimization problem is solved for surface charge density on the electrode surfaces, rather than in the full 3-dimensional space, making this part of the computation faster. The solution can then be propagated from the boundary to the region of interest. Depending on the complexity of the geometry, the performance of the COMSOL BEM and FEM varies \cite{SimonRagg2020}. 

\subsection{Radial confinement and choice of DC electrode width}\label{Sec:radialsimulations}
Once a rough geometry for the trap has been outlined, in order to simulate the radial confinement in the trap
it is necessary to select an electrode-to-ion distance, a trap drive frequency and the RF voltage applied to the electrodes. 
In the initial design phase, we targeted a zero-peak RF voltage of 370 V at a drive frequency of 2$\pi \times$90 MHz to trap both calcium and beryllium ions, an ion-electrode distance of 185 $\mu$m and an opening angle larger than 80 degrees, following a similar 3-dimensional trap operated in our laboratory \cite{Kienzler2015}. We later operated the trap with somewhat different parameters: an RF zero-peak voltage of up to 200 V at a drive frequency $\omega_{RF}$ = 2$\pi \times$36 MHz with calcium ions. Without an additional DC quadrupole to break the symmetry, these values would result in radial frequencies for a single calcium ion of $\omega _x$ = $\omega _y$ = 2$\pi \times$5 MHz.
The axial confinement is provided by applying voltages to the DC (or control) electrodes. Depending on the role of the electrodes in a specific zone of the trap, there may be a more optimal electrode width to satisfy a given requirement. An example of this is that of groups of electrodes used to split potential wells for the separation of ion chains. For 3-dimensional traps, one shuttling task which is typically limited by the applied voltages is ion separation, which requires a large quartic term. We optimise the electrode width for this parameter finding an optimal width of 160 $\mu$m (although we note that the dependence of the trapping frequency on this width is fairly weak \cite{Kienzler2015}).
The DC electrodes surrounding the junction are also important, since they are responsible for producing potential wells in the presence of the anti-confining axial fields from the pseudopotential barriers, which we will discuss in the next section. Therefore we also choose their width such that the maximal axial frequency can be produced for a specific voltage \cite{Reichle2006}. For the ion-electrode distance of 185 $\mu$m used here we found 140 $\mu$m to be the optimal width.

\subsection{Double junction bridge geometry}\label{Sec:Double junction bridge geometry}
A junction consists of two linear trap arrays intersecting with each other. There are several potential junction geometries, which we refer to through the letter corresponding to the shape of the intersection: X, T and Y. The X-junction geometry consists of two linear trap arrays intersecting at 90 degrees. In a junction in which the electrodes have mirror symmetry in three orthogonal planes which intersect the junction center, there can be no pseudopotential confinement at the center of the junction \cite{BlakestadNIST2010}. 
In order to confine ions at this location, it is therefore necessary to break the field symmetry by modifying the geometry of the electrodes in an appropriate manner. This provides pseudopotential confinement in the plane of the X, but as a result also produces axial RF electric field components away from the junction center, which we refer to henceforth as pseudopotential barriers. 

\begin{figure}[h]
\centering
\includegraphics[scale=0.65]{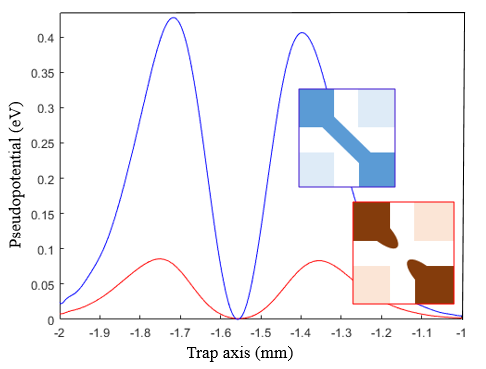}
\caption{Comparison between the RF bridge geometry (in blue) similar to that used in \cite{BlakestadNIST2010} versus an open-bridge geometry (in red) for an ion-electrode distance of 185 $\mu$m, a 2$\pi\times$90 MHz RF drive frequency, an RF voltage of 370 V and $^9$Be$^+$ ions. The height and gradient of the pseudopotential barriers can be greatly decreased using an open-bridge geometry, however this also decreases the pseudopotential confinement at the junction center. The slight imbalance in the blue geometry is due to computational noise.}
\label{fig:bridge_comparison}
\end{figure}

\begin{figure}[h]
\centering
\includegraphics[scale=0.45]{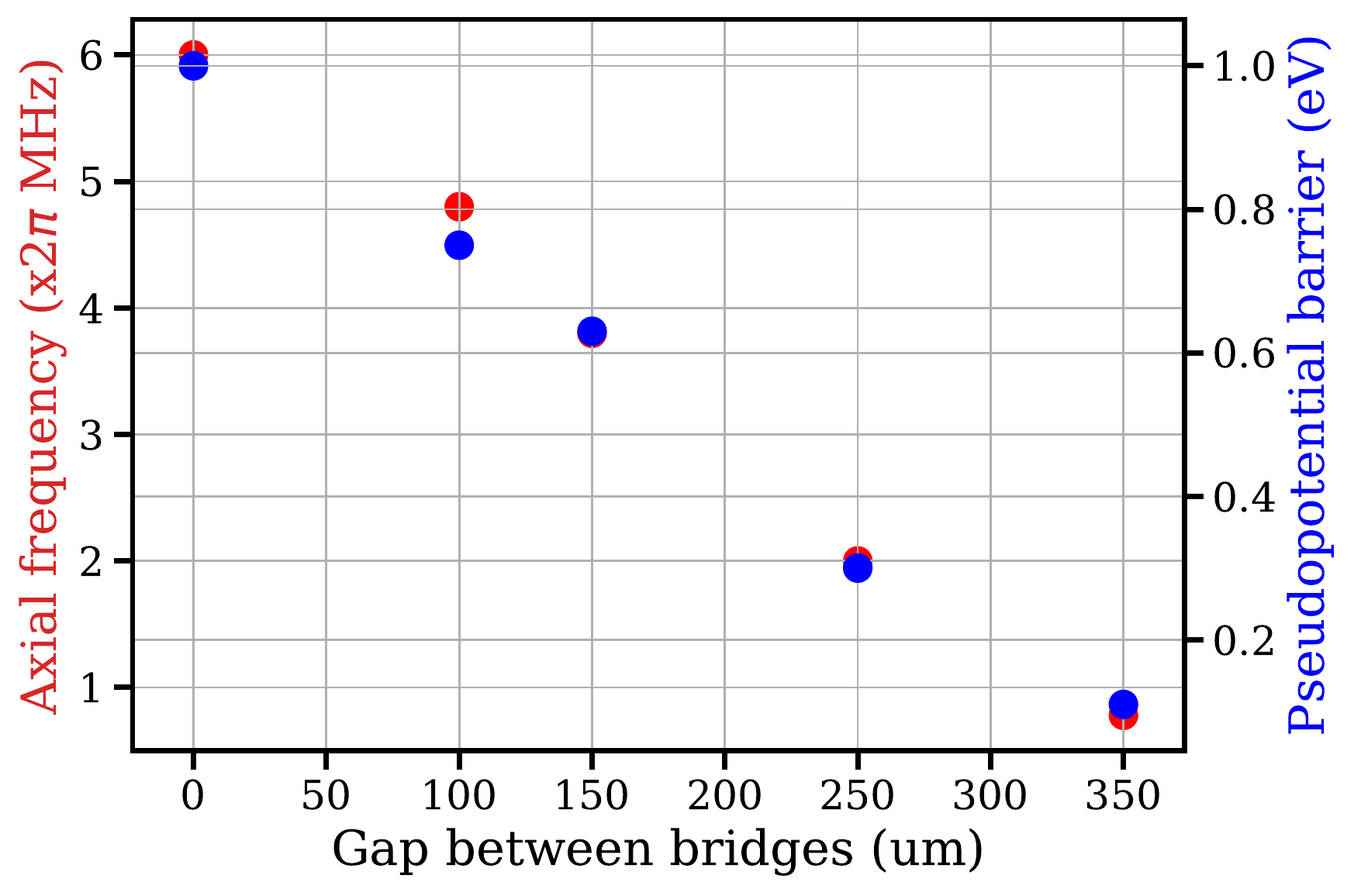}
\caption{Pseudopotential barrier height and axial frequency at the center of the junction as a function of the gap size between RF open bridges. Data simulated for the original targets of 370 V RF, 2$\pi\times$90 MHz RF drive frequency, and beryllium ions. The height and shape of the bridges is further optimised in the final design, shown in Figure \ref{fig:bridges33}. }
\label{fig:gaps_bridges_study}
\end{figure}

\begin{figure}[h]
\centering
\includegraphics[scale=0.33]{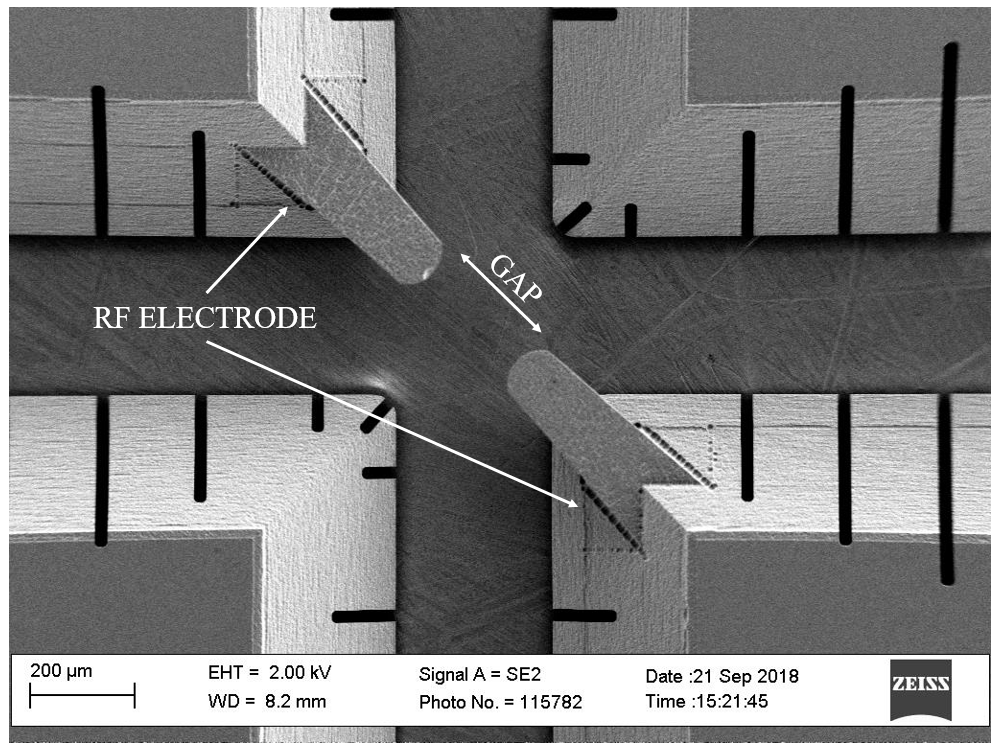}
\caption{Scanning electron microscope (SEM) picture of the chosen RF electrode open-bridge geometry for low pseudopotential barrier height and gradient.}
\label{fig:bridges33}
\end{figure}

\begin{figure}[h]
\centering
\includegraphics[scale=0.5]{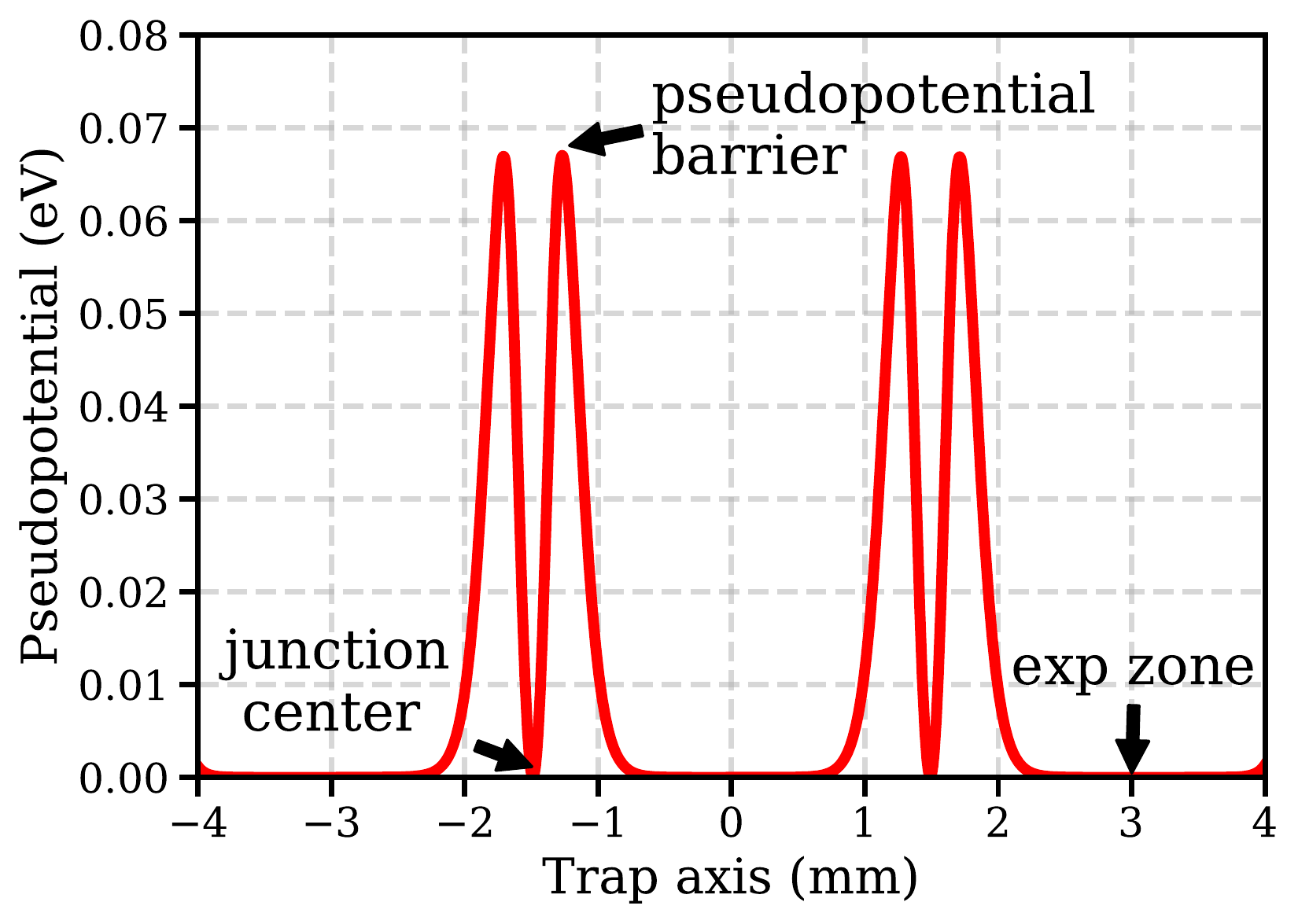}
\caption{Pseudopotential along the trap axis (z) for a beryllium ion when 370~V are applied on the RF electrode, while keeping all other electrodes grounded, at a 2$\pi\times$90~MHz RF drive frequency.}
\label{fig:axial_pseudopot}
\end{figure}

At NIST, the symmetry breaking was implemented by connecting the two opposing RF electrodes with a ``bridge'' \cite{Blakestad2009high}, in part limited by the fabrication involved. These produced large pseudopotential barriers for beryllium ions, on the level of 0.3~eV. Considering the more advanced fabrication which we used for our design, we decided to explore alternative geometries of the RF electrodes which allow for a  compromise between the height and gradient of the pseudopotential barriers and the confinement at the junction center. This also reduces the voltages required to achieve an overall confining well along the trapping axis during transport into the junction. In a reduced simulation of an isolated junction, we varied the extent of protrusions from the corner of the RF electrode (these are shown in Figure \ref{fig:bridge_comparison}, along with calculations of a pseudopotential from the geometries shown). Results of this study are shown in Figure \ref{fig:gaps_bridges_study} as a function of the gap between the protrusions. These confirm the relationship between the junction confinement and the pseudopotential barrier height.  In a similar study we found that the width and thickness of the protruding features have only a slight influence on the pseudopotential barriers. Figure \ref{fig:bridges33} shows an SEM image of the final design, with a gap of 230 $\mu$m and a 240 $\mu$m vertical distance to the center of the trap. This choice was made such that with the planned parameters of 2$\pi\times$90~MHz RF drive frequency, RF voltage of 370~V and $^9$Be$^+$ ions we would be able to attain an axial frequency at the junction center around 1 MHz. Plots of the pseudopotential on the trap axis for these settings are shown in Figure $\ref{fig:axial_pseudopot}$. Initial tests of the trap have operated at lower RF voltages and drive resulting in a junction center frequency of 0.4~MHz. 

\subsection{Residual axial fields}\label{Sec:gapsAndMisalignments}
\begin{figure}[h]
\centering
\includegraphics[scale=0.35]{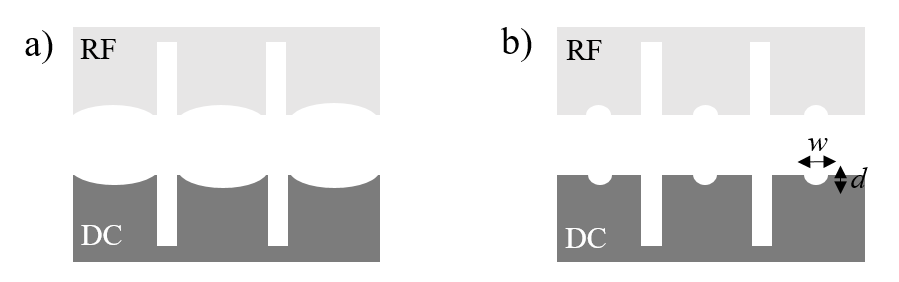}
\caption{Top view of models used for the study of shaping of electrodes with curved edges and indentations. a) curved edges electrodes: we study curvatures with depth ranging from 30 to 2~$\mu$m. b) small indentations at the centre of the electrodes edges. Results obtained with this shaping method are shown in Figure \ref{fig:shapingRF_dent_withele}.}
\label{fig:gaps_shaping}
\end{figure}

Intrinsic micromotion, which cannot be compensated with compensation fields, can arise from different geometric sources; they produce a lack of symmetry around the trap center. This can be caused by trap misalignment and fabrication inaccuracies, but also due to non-centred positioning of the ion within a finite size electrode structure \cite{Herschbach2012}. It creates a residual electric field on the trap axis which modulates the ion-laser interaction, resulting in reduced transition strength and additional undesired resonances. To understand this better for our design, we  investigated the effect of the finite size and gaps between electrodes, and of misalignment between the trap wafers. We were first interested in assessing how the residual field varies with the gap size, and whether mirrored gaps on the RF electrode help with minimising it. Furthermore we explored whether electrode shaping could be used to mitigate this effect. The following studies were conducted in a reduced linear trap made of 7 electrodes, otherwise identical in dimensions to the double junction trap (further details on the model used for these simulations and the full results are provided in the Appendix). \\
For 20 $\mu$m wide electrode gap, using simulation parameters mentioned in section \ref{Sec:Double junction bridge geometry}, we obtained electric fields along the trap axis of up to 100-200 V/m, with zero field points at axial position corresponding to the center of each electrode. We found that 40 $\mu$m-wide gaps produce axial fields 3 times as large. We decided to proceed with gaps which are 20 $\mu$m wide, which is compatible with the minimal size structure on the gold evaporation masks. We then investigated whether mirroring the segmentation of the DC electrodes on the RF electrode could help in minimising the axial field. We found this to be valuable for our setting, in agreement with similar studies \cite{Mehlstaubler2012}. We were also able to optimize the depth of the RF electrode gaps, resulting in a gap depth of 280~$\mu$m. \\
To have a reference against which to judge the effect of misalignment between the wafers, we also simulated this. We investigated two types of misalignment, a vertical angular misalignment (tilt), which could originate during the gluing stage, and a linear misalignment between the wafers, which could result from machining tolerances not being met during the substrate fabrication process, or from inhomogeneity in the metal layer thickness. A linear misalignment between
the two trapping wafers of between 10 and 20 $\mu$m was simulated. The axial electric field differs by up to 100 V/m from the one obtained in an aligned trap. However, linear misalignment at this level is unlikely to occur: tests of our alignment procedure resulted in measured misalignment of 1-2 $\mu$m \cite{simon2019}. We also investigated the effect of a tilt between the two wafers of up to 2.5 degrees. Here the residual axial field deviates by up to 100 V/m from the ideal case. This level of tilt is far above what was observed in measurements of test assemblies, which show a misalignment in the order of 0.05 degrees \cite{simon2019} - the latter would produce deviations which are in the level of noise of the simulation.\\
Finally we explored the possibility of shaping RF and DC electrodes to reduce the residual axial RF electric field. The standard geometry is that of an electrode finger with a face parallel to the trap axis. This raises the question whether a modification to the electrode shape (which becomes available given the capabilities of the laser machining that we use) can nullify the effect of the electrode gaps. The first geometry considered was a curved end along the entire width of the electrode - shown in Figure $\ref{fig:gaps_shaping}$ a). We found that curved electrodes overcompensated the axial field down to levels of curvature of 2~$\mu$m, which did have a slight compensation effect. This is at the level where the fabrication quality would be hard to achieve. 

\begin{figure}[h]
\centering
\includegraphics[scale=0.5]{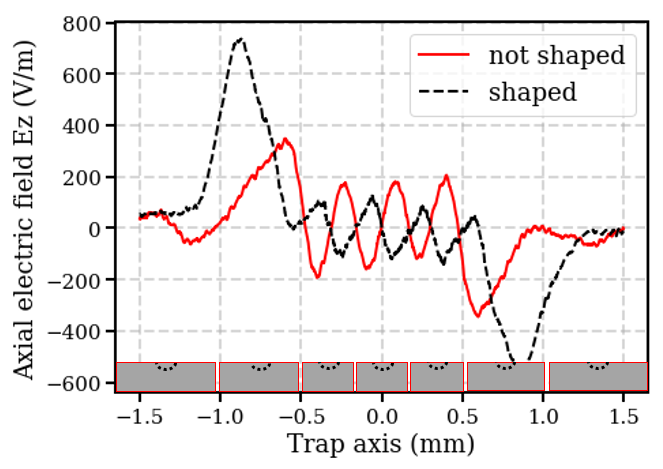}
\caption{Simulation showing the effect of shaping electrodes in order to minimise $E_z$. Electrodes are shaped as shown at the bottom of the figure, with small curved indentation 60 $\mu$m wide and 5 $\mu$m deep. A non-shaped electrode is a standard straight face geometry, while the shaped electrodes have a small indentation as shown in Figure $\ref{fig:gaps_shaping}$ b). Details on the simulation model are provided in the Appendix.}
\label{fig:shapingRF_dent_withele}
\end{figure}

We then studied a different shaping method. Instead of curving the whole finger edge, we created an indentation in the middle of the finger, as shown in Figure \ref{fig:gaps_shaping} b) (and repeated in the inset at the bottom of Figure \ref{fig:shapingRF_dent_withele}). We find that for a 60 $\mu$m-wide ($w$) and 5 $\mu$m-deep ($d$) indentation the sign of the electric field is flipped, with the amplitude of the deviations from zero being slightly reduced. This can be seen in Figure $\ref{fig:shapingRF_dent_withele}$. We therefore think that this could be further optimized. However we decided not to take the risk of additional features, since the fabrication is already quite demanding. The small fields required to cancel out those produced by the gaps can be achieved only with very small features, which could easily be distorted during the fabrication process. Such a study, however, is insightful in providing an estimate for the fields generated by variations of the electrode surfaces in the range of a few micrometers in size, which could arise either due to the machining stage, as a result of mishandling of the wafers, as discussed in \cite{simon2019} or during metal deposition e.g. electroplating inhomogeneity. Such defects, in traps similar to the one described here, may produce residual axial fields of ten to a few hundred volts per meter.

\section{Trap fabrication and packaging}\label{sec:fabrication}
\begin{figure}[h]
\centering
\includegraphics[scale=0.42]{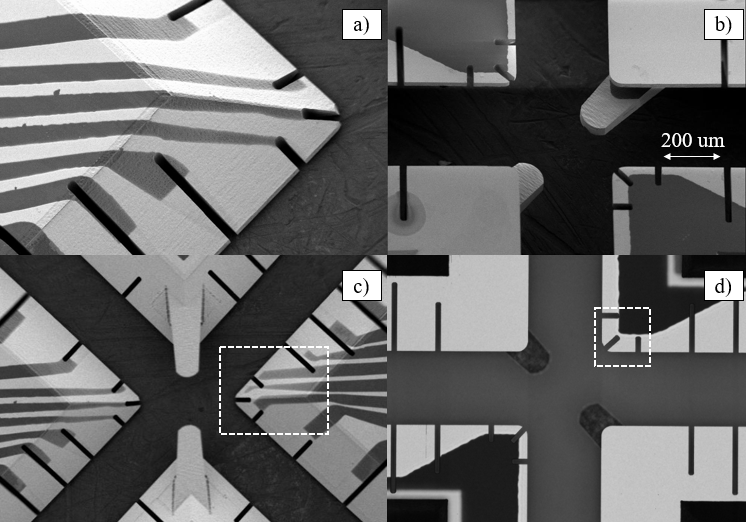}
\caption{SEM pictures showing gold evaporation using shadow masks on the top and bottom side of a single trap wafer. Insets a) and b) show a successful evaporation, while in c) and d) a misalignment of the masks resulted in shorts between electrodes, highlighted by the white dotted boxes.}
\label{fig:evap_goodandbad}
\end{figure}

\begin{figure}[h]
\centering
\includegraphics[scale=0.27]{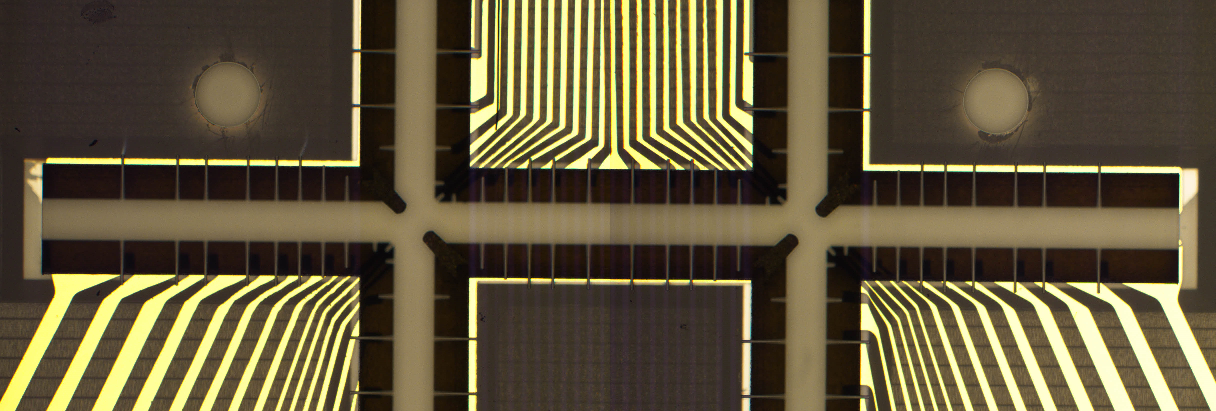}
\caption{Top view of one of the trapping wafers, showing all the DC electrode traces evaporated using several shadow masks.}
\label{fig:full trap}
\end{figure}

The laser machining of the trap wafers is performed by FEMTOPrint \cite{FEMTOprintWeb}. Pior to gold coating, we implement a cleaning step. Due to observed fragility of the glass wafers, we opt for a chemical cleaning process instead of an ultrasonic cleaning process. After a Piranha cleaning step, electron-beam evaporation at different tilt angles is used to deposit a uniform layer of 200 nm of titanium (as the adhesion layer) and 750 nm of gold, as previously done \cite{BlakestadNIST2010, Kienzler2015}. This creates a total metal layer of 950 nm. We do not electroplate the sample due to concerns that electroplating might introduce a higher surface roughness and in-homogeneity in thickness across the electrodes at the level of a few micrometers, which would produce residual axial fields as we discussed in the previous section. Due to the presence of 3-dimensional features, several evaporation rounds are required at different tilt angles to ensure coating of all surfaces. \\
The electrodes are defined using a number of custom-made laser-cut metal masks. We use 100 $\mu$m-thick Molybdenum masks. The smallest electrode width is in the order of 10 $\mu$m. For areas with a high density of electrodes (around the DC junction electrodes) multiple optimisation rounds have been carried out to obtain a mask with the required precision. Details on the laser cutting process can be found in \cite{NorbertAckerl2020}. The masks are manually aligned to better than 5 $\mu$m precision under a microsope and clamped to the wafer before evaporation. Due to the 3-dimensional features of this trap, the mask alignment is very challenging, and several trials are required in order to avoid shorts between electrodes. Figure $\ref{fig:evap_goodandbad}$ insets a) and b) show a successful evaporation on the top and bottom side of one of the trap wafers, while insets c) and d) show unsuccessful evaporation rounds where a misalignment created shorts between the electrode traces. Figure $\ref{fig:full trap}$ shows all of the DC electrodes for the top wafer. These traces were evaporated using 4 separate masks.

\begin{figure}[h]
\centering
\includegraphics[scale=0.4]{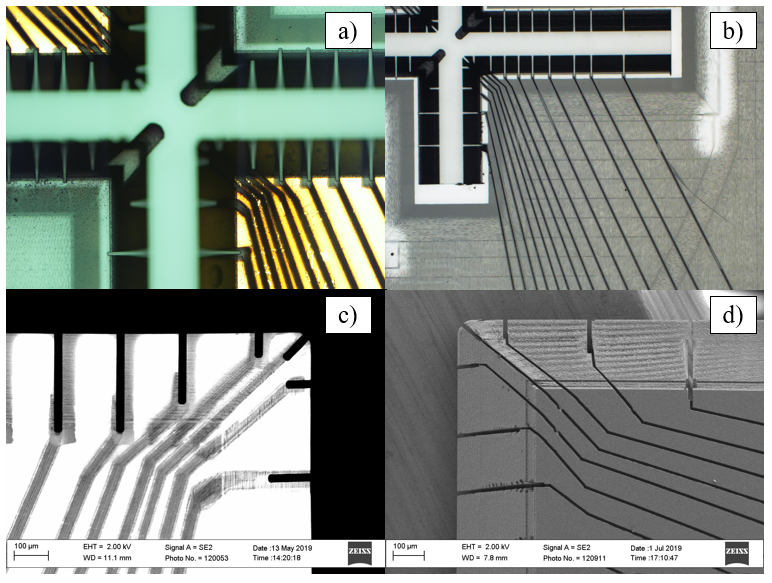}
\caption{Early fabrication tests using ablation and trenches. After gold evaporation, a laser is used to ablate material between electrodes. On the other hand, trenches can be machined on the wafer, before gold evaporation. a) microscope picture of ablated sample. b) microscope picture of laser-machined trenches. c) SEM picture of ablated sample in the region of highest electrode density. d) SEM picture of laser-machined trenches in the region of highest electrode density. }
\label{fig:trenches_ablation}
\end{figure}

Due to the challenge in producing masks with very fine features and aligning them with enough precision on the substrate wafers, we have investigated two alternative fabrication methods: the first consists of defining the electrodes by ablating the metal after an evaporation step, the second involves incorporating trenches in the wafer design itself to isolate the electrodes, as done in \cite{panoptic2020}. Particularly promising is the use of trenches inbuilt in the wafers which eliminates the need for multiple and complex shadow masks, and avoids the risk of shorting electrodes due to misalignment of the masks. However, in order to ensure wafer stability, trenches should be implemented in thick wafers (more than 300 $\mu$m thick). Figure \ref{fig:trenches_ablation} shows initial ablation and trench tests. While the ablation process achieves a slightly worse result than our laser machined and manually aligned masks (Figure \ref{fig:trenches_ablation} panels a) and c)), the trenches can have a smaller width and therefore are particularly suitable for very dense electrode areas such as our junction electrodes (Figure \ref{fig:trenches_ablation} panels b) and d)).

\begin{figure}[h]
\centering
\includegraphics[scale=0.35]{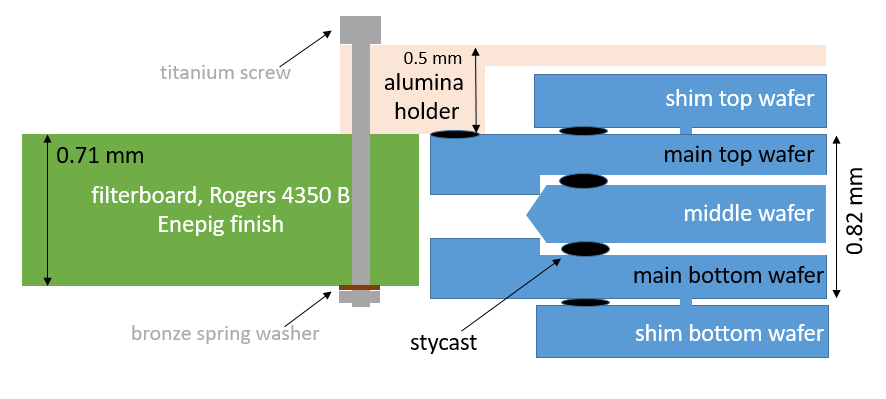}
\caption{Side view of the trap assembly, including the gold-coated alumina wafer screwed to the filterboard. Droplets of stycast are used to fix the different trap wafers.}
\label{fig:assembly_final}
\end{figure}

\begin{figure}[h]
\centering
\includegraphics[scale=0.7]{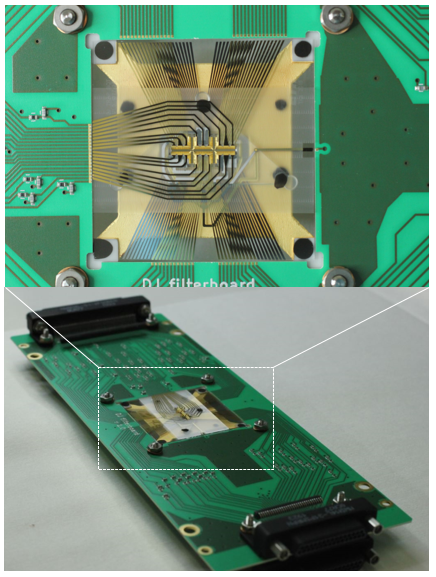}
\caption{Top: assembled trap fixed to the filterboard. Black dots on the wafers are the stycast droplets used to glue the wafers, the droplets are located at carefully chosen alignment locations. Bottom: filterboard PCB with gold-coated alumina wafer and trap attached in the center pocket through screws. Micro sub-D connectors are used at the edges of the filterboard to connect all DC lines.}
\label{fig:trap_and_PCB}
\end{figure}

In order to host all 143 electrodes, it was necessary to route the RF electrode on the opposite side of each wafer than the DC electrodes. A wafer via allowed to bring the RF electrode back on the outer wafer side to be wirebonded. The via we implemented has a diameter of 300 $\mu$m and a depth of 300 $\mu$m, as previously discussed in \cite{simon2019}. Electrical tests showed an increase in the track resistance on the order of 1-2 Ohm across the via. To ensure proper operation of the trap, the RF electrode on the top and bottom wafer need to have the same phase of RF signal. A difference in track length would result in a phase mismatch between the top and bottom wafer RF. In order to avoid this, the RF tracks hosted on the top and bottom wafer respectively are matched in length and width. In our case, the width of the RF tracks is 300 $\mu$m throughout. On the PCB, the RF track is surrounded by a ground plane. Once the electrode tracks are defined on each individual wafer, the wafers are aligned to each other using self-aligning guides incorporated on the wafers \cite{simon2019} and glued in position using Stycast. We cure the glue at room temperature for 16 hours. The trap capacitance including the PCB to which the trap is attached is measured to be 14 pF.\\
For the packaging of the trap, we considered having an intermediate PCB to which the trap would be wirebonded, connected to a larger PCB with DC filters. However, due to constraints on the size of the PCB and the trap, and the large number of connections, we decided to wirebond the trap directly to the filter PCB. The trap stack is first glued to a gold-coated alumina holder, which is screwed to the filterboard and connected to ground, as shown in Figures $\ref{fig:assembly_final}$ and $\ref{fig:trap_and_PCB}$. In order to maintain flexibility in the assembly without over-constraining it, we screwed the alumina holder to the filterboard using bronze elastic washers which ensure appropriate tightness when cold. 

\begin{figure}[h]
\centering
\includegraphics[scale=0.3]{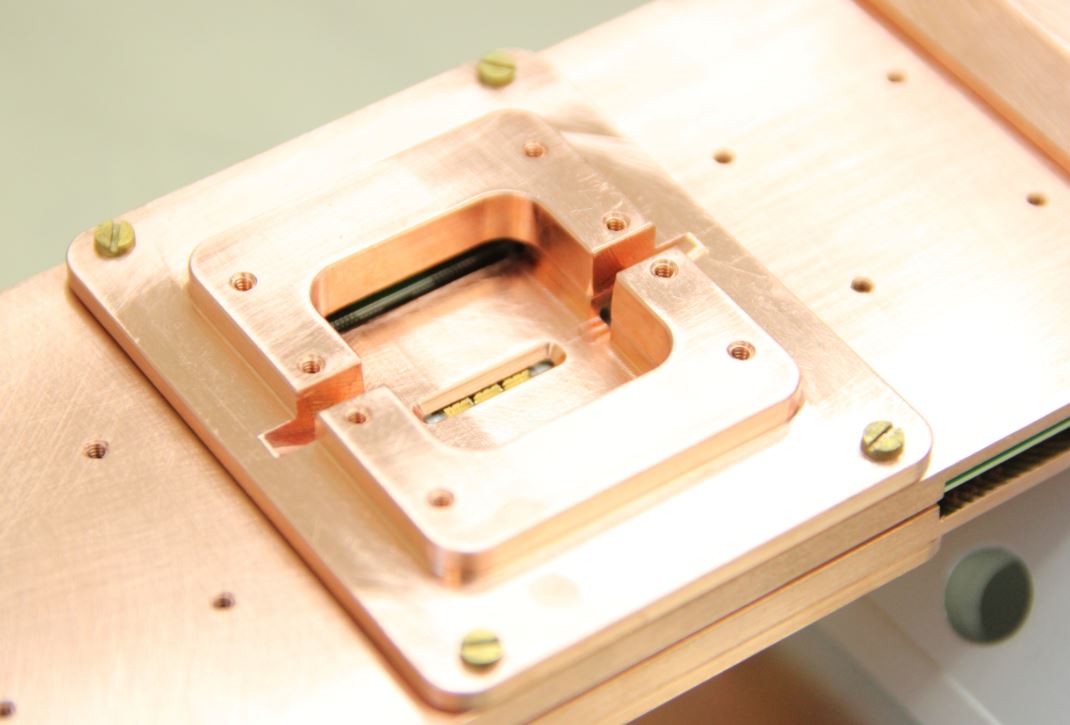}
\caption{A copper sleeve surrounds the assembled trap and filterboard. An additional shield with a narrow opening is placed right above the trap and protects the wafer from neutral calcium atoms, while also providing a ground plane.}
\label{fig:copper_sleeve_light}
\end{figure}

We use a 700 $\mu$m thick filterboard made of Rogers 4350 B for low RF loss. The primary role for this filterboard is to provide a ground to the RF signal. An initial filtering stage for the DC lines is supplied by first-order RC filters with 240 Ohm resistors and 820 pF capacitors, resulting in a 809 kHz corner frequency. Micro sub-D connectors are used to connect the DC signal lines on the filterboard to Kapton-insulated manganin wires. The wires are home-made and about 60 cm long, they connect the DC lines on the trap filterboard to additional filter-PCBs external to the cryostat. Due to the high number of DC lines, we populate the filterboard PCB on both sides, and use pairs of Micro sub-D connectors (one of the top side and one on the bottom side) at each edge of the filterboard. The whole trap and PCB assembly is shown in Figure $\ref{fig:trap_and_PCB}$. The filterboard is inserted in a copper sleeve which serves as a mounting piece and a protective layer and includes an oven shield (Figure $\ref{fig:copper_sleeve_light}$). 
 
\section{Experimental trap characterisation}\label{sec:measurements}
The trap is inserted in a low-vibration cryostat and cooled down close to 4\;K. The RF signal is provided by a home-made helical RF resonator (size 6$\times$4$\times$4 cm) placed close to the trap in the 4\;K chamber, similar in design to \cite{SilvernsResonator}. We design an antenna tuning mechanism which allows reliable tuning of the antenna coil to match the main coil inductor, however the impedance matching of the resonator can only be performed upon warming up the cryostat and opening the inner 4\;K chamber. We use a diffusive-type oven as a source of neutral calcium atoms which are photo-ionised \cite{photoionisation2004}.

\subsection{Radial heating rate}\label{Sec:radial heating}
Laser-cooled trapped ions are a very sensitive probe of electric fields. 
As a result, electric field noise is an important parameter affecting multi-ion coherent control, since it heats the ions producing decoherence of shared oscillation modes \cite{brownnutt15}. By measuring the heating rate $\Gamma_h$, we can determine the spectral density $S_E(\omega_m)$ of electric-field noise at the ion's motional frequency following \cite{turchette00}:

\begin{equation}
    \Gamma_h = \frac{e^2}{4 m \hbar \omega_m} S_E(\omega_m)
\end{equation}

There are several methods to measure the temperature of a trapped ion \cite{brownnutt15}, which when measured as a function of a delay time yields the heating rate. For measuring heating rates of the radial oscillations we use a tightly focused beam orthogonal to the axis. We start by using resolved sideband cooling to cool the ion close to the motional ground state \cite{Wineland1997experimental}. The thermal occupancy is then estimated using sideband thermometry, which consists of probing the first red and blue motional sideband with a fixed duration probe and comparing the excitation probability \cite{monroe95,turchette00}. In some measurements, we monitored the time evolution of the internal state as a function of time of a blue-sideband or carrier probe pulse. Fits of expected curves with thermal motional population distributions were then used to extract the mean oscillator occupancy. 

\begin{figure}[h]
\centering
\includegraphics[scale=0.5]{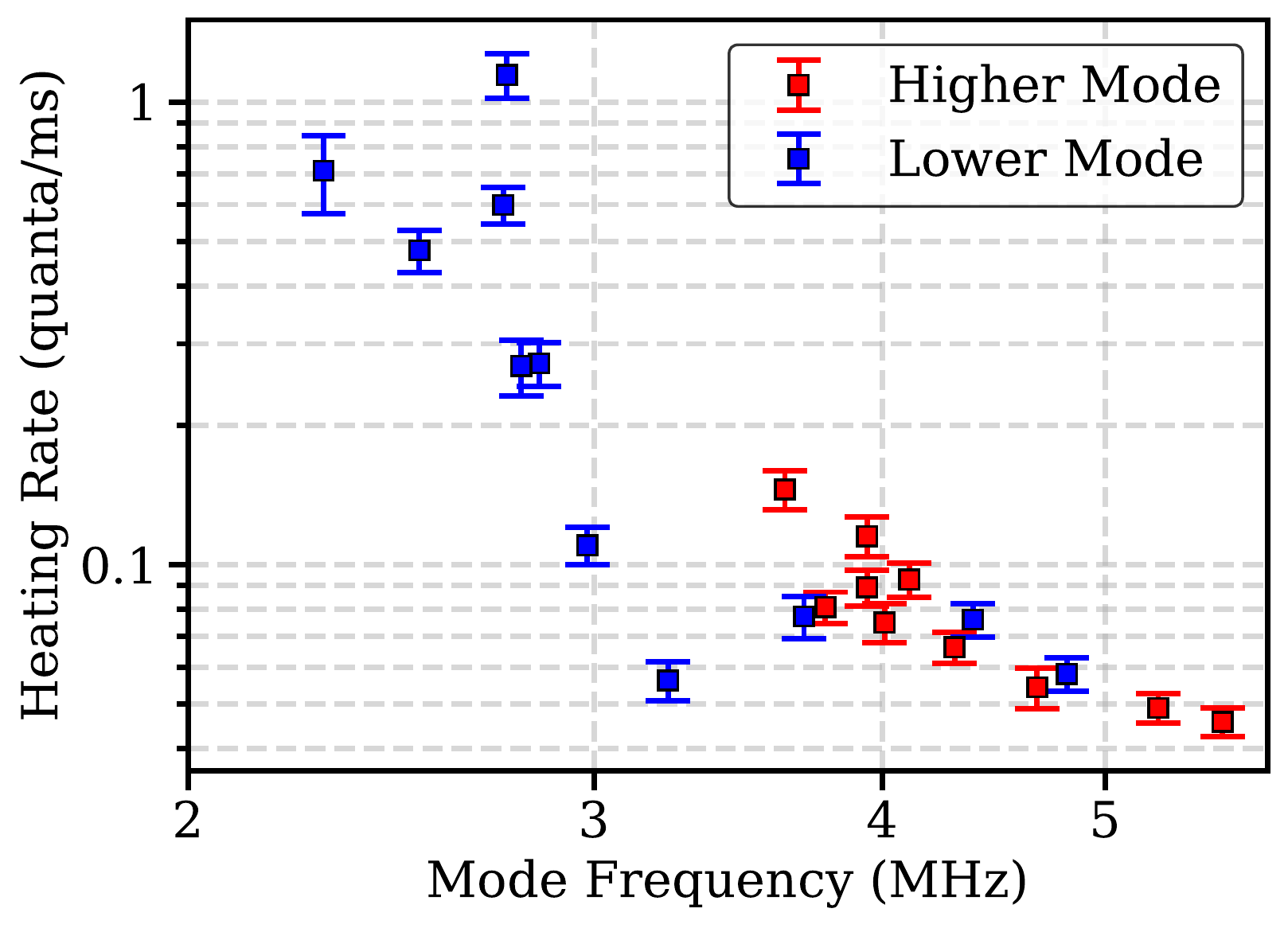}
\caption{Radial heating rates of both the high (red) and lower (blue) frequency mode after cooling to the quantum ground state.
}
\label{fig:Radial heating}
\end{figure}

In first measurements we obtained a radial mode heating rate close to 1500 quanta/s at 4.4\;MHz. The noise was found to be strongly polarized along the direction between the RF electrodes, which is surprising since these are driven from the same RF supply, with the signals to the two electrodes divided on the wafer. It was observed that the rate did not depend strongly on the excess micromotion compensation, which suggests that it did not originate in the RF part of the signal. \\
After making a number of changes to the RF drive setup, the heating rate was lowered to around 80 quanta/s, a reduction of about 20 times, and lost its polarisation. The main modifications to the apparatus that were implemented between these two measurements were the following 1) a rectifier providing a monitor of the RF drive after the helical resonator in the 4K chamber was removed, 2) an additional copper shield was installed on the side of the ion trap, to provide a more symmetric grounding environment, 3) high resistance RF cables (UT-085-SS-SS from Amawave) from the 300 K to the 4 K stage were substituted with low-resistance cables (UT-085TP) and 4) nano-positioner cables, used to move the in-cryo imaging objective, were routed further away from the RF line. \\
Following these experimental upgrades, we measured the radial heating rate as a function of RF input voltage. Figure \ref{fig:Radial heating} show the radial heating for the two radial modes. We observe that the radial heating increases dramatically as we reduce radial secular frequencies below 3 MHz and there is a peak of heating at around 2.75 MHz. A radial heating rate around 80-100 quanta/s still seems high for a microfabricated ion trap, similar to \cite{BlakestadNIST2010} and \cite{Kienzler2015}. Both those traps, with comparable geometry and ion-to-electrode distance, measure heating rate well below the one measured in this trap, and furthermore were operated at room temperature.

While it is hard to identify the exact sources of noise, we suspect the high heating rate is due to residual technical noise, coupled for instance through the long DC-line cables connecting the room-temperature filters and the trap PCB.

\subsection{Axial heating rate}\label{Sec:Axial heating}
We measure the axial heating rate via a different technique, because due to a combination of high observed heating rates and a lack of laser intensity in the laser beam which is able to couple to the axial modes we can not cool the axial mode close to the motional ground state. Thus we instead first Doppler-cool the ion, and then drive Rabi oscillations on the carrier transition with a laser at 45 degrees to the axis. We fit these with known expressions from thermal states to obtain the mean occupation of the axial mode. Since the radial heating rates are much lower than that of the axial modes, we assume that the axial dominates the motional effect on the Rabi oscillations. \\
The axial heating rate was measured over a range of different axial frequencies between 500~kHz and 2.2~MHz. To investigate whether this heating rate depends on the temperature of the trap chip, we  measured the heating at a fixed axial frequency with two settings for the RF drive which differ in power to the trap by a factor of 2. This had no noticeable effect on the axial heating rate. Although the source of this noise remains to be found, we think that the temperature insensitivity suggests that axial heating is from technical sources. Implementing setup modifications such as disconnecting shim electrodes, secondary filters, oven lines, and disconnecting the chamber ground from the optical table did not have any effect on the measured axial heating rate. One suspicion is noise picked up by the relatively long (60 cm long, to reduce heat conductivity) resistive cables between the room-temperature and cryo-stage of the experiment, which are then only filtered by the cryogenic filter board with an 800~kHz cut-off frequency. 

\begin{figure}[h]
\centering
\includegraphics[scale=0.5]{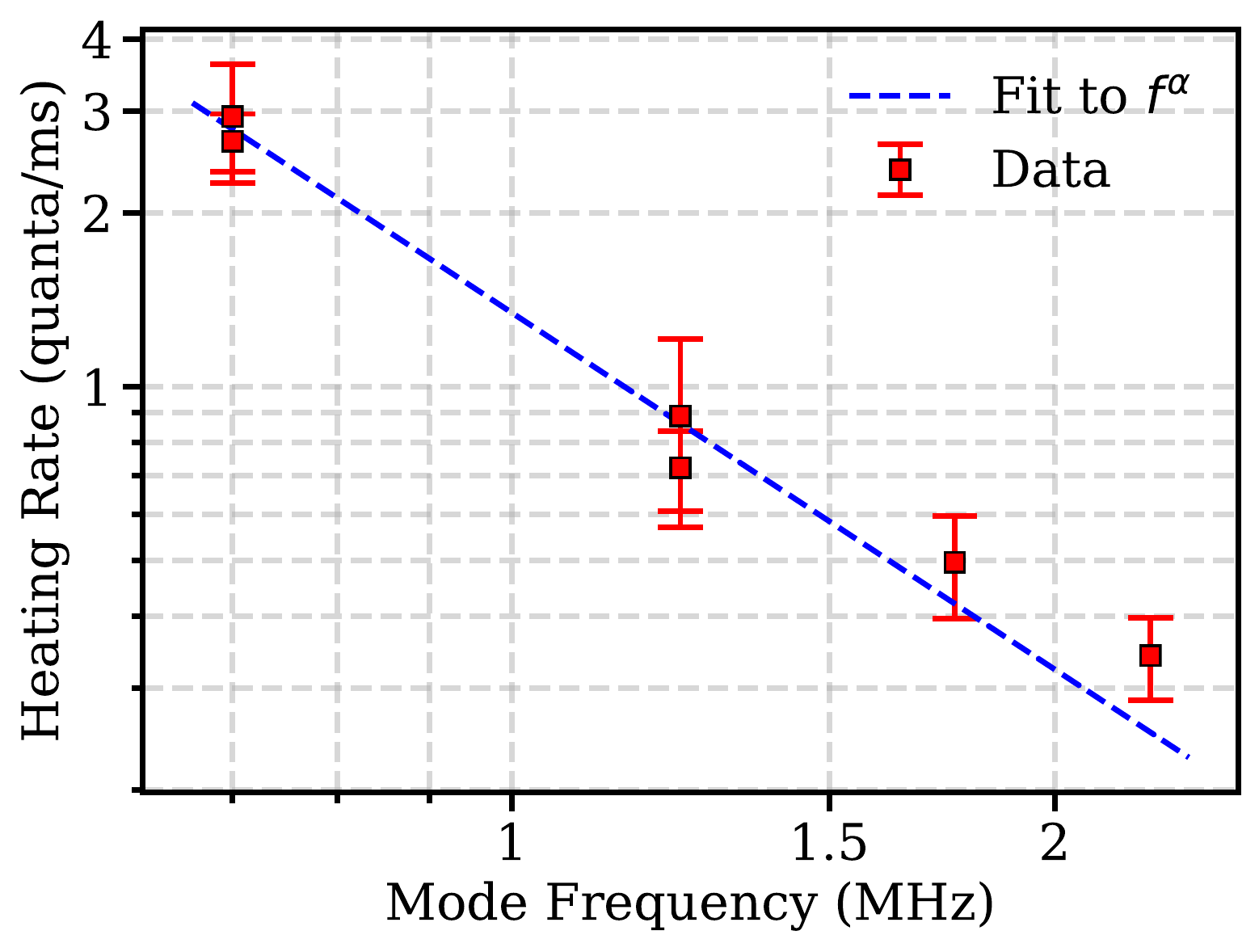}
\caption{Axial heating rate measured via measuring slow-down on Rabi oscillations for different wait times while driving the carrier transition. Axial heating rate fitted to f$^\alpha$. (Power law fit with $\alpha = -2.05$)
}
\label{fig:axial heating}
\end{figure}

\subsection{Axial micromotion}\label{Sec:Axial micromotion}

In order to characterize axial micromotion, we compare the Rabi frequency while driving the first micromotion modulation sideband and the central feature of a quadrupole transition at 729~nm, using the laser beam propagating at 45 degrees to the trap axis. The ratio of the two allows us to estimate the modulation index, from which we can extract the electric field on the axis of the trap. We control for the radial modulation of the laser due to excess micromotion by minimizing this using a modulation technique \cite{brownnutt15}. \\
With this method we are able to suppress the first micromotion sideband for a radial 729~nm laser beam by a factor of 100.  For the 45 degree beam, we then find that the sideband-to-carrier ratio is 1:40. This allows us to obtain a modulation index $\beta$ = 0.05 for the axial micromotion. We convert this to an electric field using:
\begin{equation}
    E_{z,\rm{RF}} = \frac{m \Omega_{\rm{RF}}^2}{k Q}\beta
\end{equation}
Where $E_{z,{\rm RF}}$ is the axial field from the RF electrode, $m$ is the ion mass, $\Omega_{\rm RF}$ is the RF drive frequency, in this case 2$\pi\times$36.3 MHz, $k$ is the wave vector and $Q$ the electron's charge. We calculate  $E_{z,RF}$=178 V/m from the measured modulation index. From our FEM simulation, which considers an ideal, perfectly aligned trap, we expect an axial RF field of around 50 V/m at our experiment location. The discrepancy could be accounted for by slight simulation inaccuracies, and by a small misalignment during fabrication. Our measured value is in agreement with similar trapping conditions \cite{Keller2015}. \\
We can compare the measured modulation index with that of a similar segmented linear trap described in \cite{Kienzler2015}. This trap has the same electrode-to-ion distance as the one described here, and a similar fabrication recipe. The modulation index measured for a beryllium ion after optimisation was $\beta$=3 \cite{negnevisky2018}. We can rescale this quantity by the ratio of the masses of beryllium and calcium to obtain the modulation index for calcium $\beta$=0.67, which is about 13 times larger than the one measured in our trap. This comparison indicates that the double-junction trap has less residual axial field. This would be expected due to the better alignment between wafers and manufacturing precision than previous manually-aligned traps. 

\subsection{Junction transport}\label{Sec:Junction transport}
We perform an initial characterisation of transport through one of the two junctions. Due to mask misalignment during fabrication, at one of the two junctions some of the DC electrodes are shorted together. These limit our voltage sets near the junction center. Nevertheless, by taking account of this in the simulation and generation of voltages for the trap, we can still perform transport despite their presence. 
\begin{figure}[h]
\centering
\includegraphics[scale=0.3]{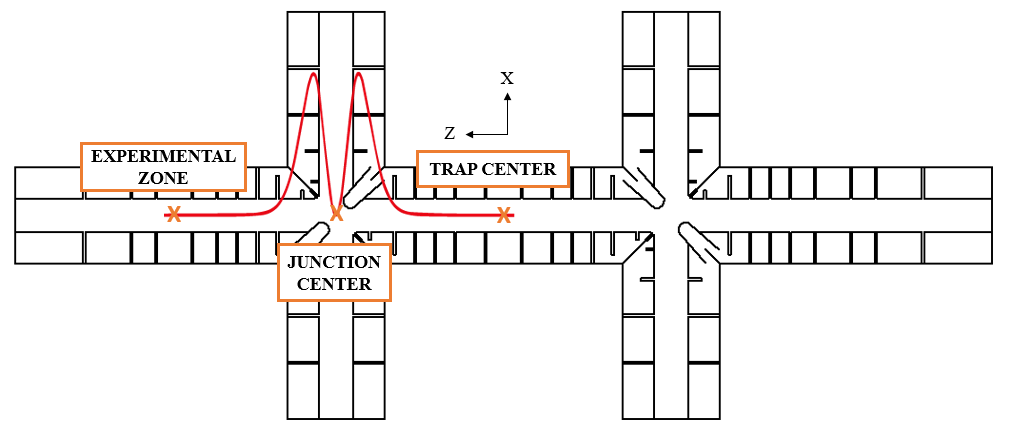}
\caption{Sketch of the left junction of the trap, with landmark locations used as starting and ending points for the transport sequences. The axial RF pseudopotential along the trap axis ($z$) is shown in red.}
\label{fig:transport areas}
\end{figure}

Transport waveforms are generated by calculating the voltage sets required to produce target potentials at various points along the trap axis ($z$-axis, as shown in Figure \ref{fig:transport areas}) and along the storage leg axis ($x$-axis, as shown in Figure \ref{fig:transport areas}). We generate a set of voltages meeting constraints of a desired set of curvatures for points spaced by 10 $\mu$m along the transport direction. The optimizer uses the same voltages for the DC electrodes on each side of the trap axis $z$. We also constrain the variation of voltages between subsequent transport steps to minimize the slew-rate requirements on the voltage supplies, and we add a regularization term that pulls irrelevant distant electrodes to 0 V. These constraints allow us to maintain symmetry and to achieve modest total voltages applied to DC electrodes during the transport sequence. Figure \ref{fig:transport areas} shows the three main locations considered in transport tasks to date, which are our starting and ending point in the transport sequences, as well as indicating the positions of the axial pseudopotential barriers. 
\begin{figure}[h]
\centering
\includegraphics[scale=0.5]{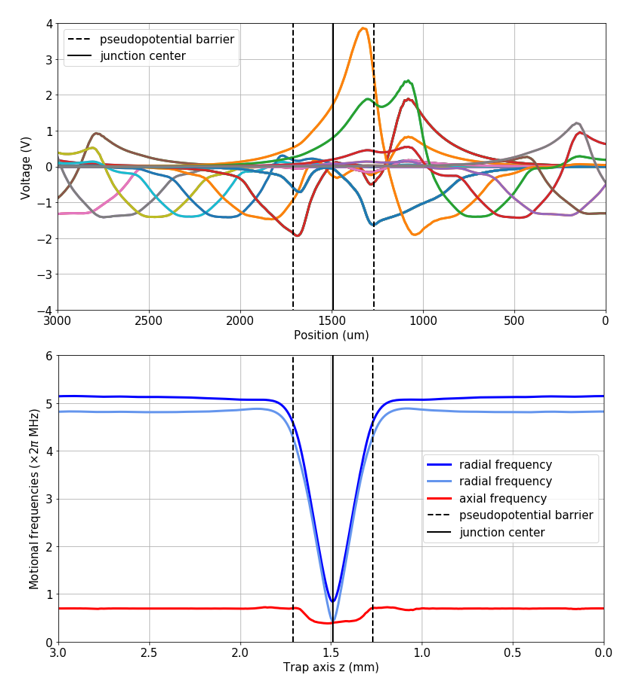}
\caption{Top: optimised voltages over position/time used to transport the ion from the experimental zone to the trap center. Note that around the junction center, shown by the black solid line at 1490 $\mu$m, the voltages are more asymmetric due to the existing electrode shorts. Modest voltages up to 4 V are required for the transport sequence. Bottom: radial and axial motional frequencies during the same transport sequence. The radial and axial frequencies decrease as we approach the junction center. At the junction center, one of the radial frequencies (along the $x$ axis), is close to the axial frequency (along the $z$ axis) due to the symmetry of the RF electrode.}
\label{fig:transport_wavef_freq2}
\end{figure}

Figure \ref{fig:transport_wavef_freq2} shows the optimised waveforms for a transport sequence starting at the experimental zone and ending at the trap center. Note that modest voltages are required in this case compared to previous realisations \cite{BlakestadNIST2010}, which in part results from the pseudopotential barriers having reduced gradient and curvature. Figure \ref{fig:transport_wavef_freq2} also shows the simulated motional frequencies during the transport sequence. All motional frequencies decrease towards the center of the junction. At the center of the junction, most of the confinement is provided by the RF pseudopotential barriers along the z and x-axis, and at this location one of the radial frequencies (along the x-axis, i.e. the storage legs) becomes near degenerate to the axial frequency. At the center of the junction the lowest motional frequencies are around 2$\pi \times$0.4 MHz. The location at which the DC electrode voltages are highest is at the point where the ion passes the apex of the pseudopotential barriers, as highlighted by the vertical dashed lines in Figure \ref{fig:transport_wavef_freq2}. Here the RF produces an axial anti-confinement, which needs to be compensated by the DC electrodes to create an overall axially confining potential. Our initial tests show that we can shuttle ions reliably from both the experimental zone and the trap center to the center of the junction (hundreds of repetitions without ion loss). 

\section{Conclusions}\label{Sec:Conclusions}
We have described the design, fabrication and experimental characterisation of a double-junction 3-dimensional trap made from silica glass wafers. The primary challenge of the fabrication was the high number of lines that need to be incorporated on the trap. Two methods might be used to simplify this process. The first involves laser machining trenches into the glass structures prior to metalization, which allows to isolate electrodes directly on the wafers rather than using shadow masks \cite{panoptic2020}. We have also experimented with ablation of gold from pre-coated samples, but the probability of shorts between electrodes seems to increase in this approach.  Initial characterisations of the trap show satisfactory levels of performance, but the heating rates are still much higher than comparable traps, especially given that this trap is being operated close to 4~K \cite{Kienzler2015, BlakestadNIST2010}. We think that this is likely due to technical noise. We find the axial micromotion to be comparable to segmented traps made for metrology applications, and significantly smaller than values reported for multi-segment alumina traps \cite{Kienzler2015} operated in our laboratory. This provides a starting point for investigations of algorithms including extensive re-configuration of ion chains in the quantum CCD architecture for large-scale quantum computing.

\section{Author contributions}\label{Sec:Contributions}
CD designed, fabricated and assembled the trap with significant contributions from SR, ME conducted early fabrication feasibility studies under CD's supervision, RM and RO built the cryogenic setup and performed experiments with CD and JF. JPH conceived the project. CD wrote the manuscript with inputs from all authors.

\section{Acknowledgments}\label{Sec:Acknowledgements}
We thank Andrea Lovera at Femtoprint for many useful discussions and the staff at the FIRST cleanroom laboratory at ETH for helpful advice on fabrication. This work would have not been possible without a tight collaboration with Norbert Ackerl (D-MATV, ETH), who developed a method to laser cut the delicate shadow masks used for gold evaporation. We thank Maciej Malinowski for helpful discussions and comments on the manuscript. We acknowledge funding from the Swiss National Fund under grant numbers $200020\_165555$ and $200020\_179147$, and from the EU Quantum Flagship H2020-FETFLAG-2018-03 under Grant Agreement no. 820495 AQTION, as well as from Intelligence Advanced Research Projects Activity (IARPA), via the US Army Research Office grant W911NF-16-1-0070.

\bibliography{BibFile}

\onecolumngrid
\newpage

\appendix
\section{Study of residual axial fields}
\begin{figure}[h]
\centering
\includegraphics[scale=0.5]{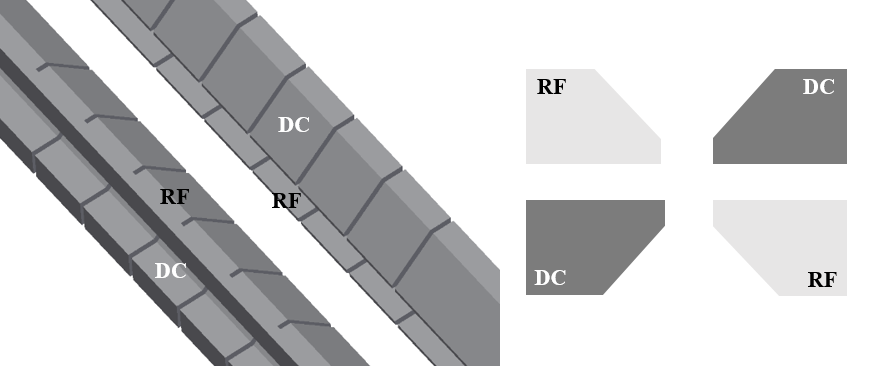}
\caption{Left: trap model used for the simulations of gaps and misalignment. We simulate the trap in a Paul configuration as shown on the right, with the DC electrodes set to ground.}
\label{fig:appendix_model}
\end{figure}

In this appendix we give details on the study of electrode gaps and misalignment, and the shaping of electrodes. As previously stated, we use a reduced linear trap made up of 7 segments for this investigation, Figure \ref{fig:appendix_model} shows the trap geometry used. The electrodes at the outer edges are longer than the inner ones, in order to minimise the effects of the ends of the trap. The boundary conditions are defined by a tight grounded airbox which surrounds the trap. A similar study has been presented in \cite{Mehlstaubler2012, Burgermeister2019}. We begin the investigation considering the residual electric field along the trap axis, with gaps present only on the DC electrodes (which are held at ground).

We study gaps on DC electrodes varying the width $w$ from 20 $\mu$m to 40 $\mu$m with a fixed depth $d$ of 300 $\mu$m. These studies show that for our geometry DC gaps which are 20 $\mu$m wide are contributing the lower residual axial field, and wider gaps create higher axial fields. We then mirror these gaps on the RF electrode and study the influence of their depth $d$. Varying $d$ from 100 to 250 $\mu$m shows that gaps between 200 and 250 $\mu$m on the RF electrode are able to minimise the axial field the most. Our complex trap geometry allows us to choose a gap depth only further away from the junction, where the density of electrodes constrains this parameter.

\begin{figure}[h]
\centering
\includegraphics[scale=0.6]{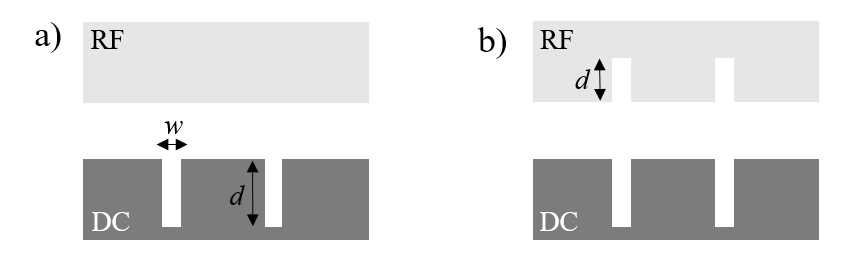}
\caption{Top view of a single trap wafer illustrating the parameters studied. a) we start by studying the residual axial field as a function of width $w$ and depth $d$ of DC electrodes gaps. b) we mirror the DC electrodes gaps on the RF electrode and vary the depth $d$.}
\label{fig:gaps_study}
\end{figure}

\begin{figure}[h]
\centering
\includegraphics[scale=0.8]{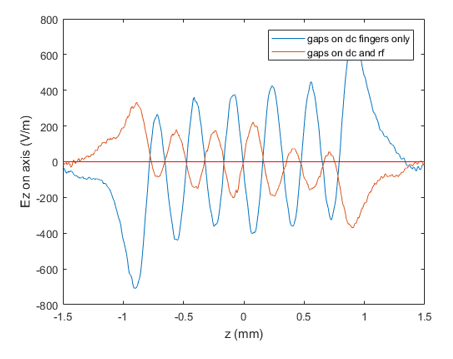}
\caption{Simulation based on the trap model presented earlier showing the effect of introducing gaps on the RF electrode which mirror those on the DC electrode.}
\label{fig:gaps_dc_rf}
\end{figure}

We compare a geometry with gaps on the RF electrode as well as on the DC electrodes and without. The gaps on the RF electrode indeed act towards canceling out the axial fields generated by the opposite gaps. The depth of the RF gaps can be modified as we described earlier to find the most helpful geometry. Figure $\ref{fig:gaps_dc_rf}$ illustrates how the axial field $E_z$ varies in the two scenarios.

\begin{figure}[h]
\centering
\includegraphics[scale=0.4]{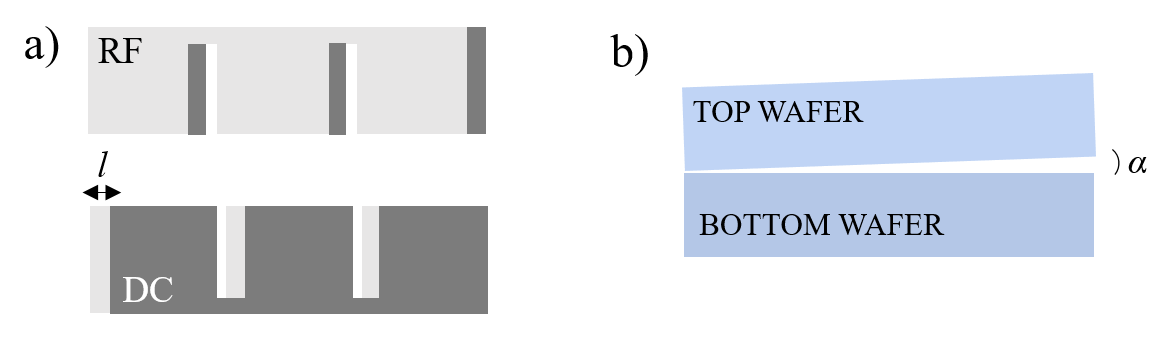}
\caption{Visualisation of the two types of misalignment investigated. a) a linear misalignment between the wafers. b) an angular misalignment between top and bottom wafer. Note that the wafers here are depicted in side view.}
\label{fig:misalign}
\end{figure}

Next we are interested in assessing the effect of misalignment between the trap wafers which we cannot correct once the trap is assembled. Figure $\ref{fig:misalign}$ shows the two types of misalignment that we study: a linear and an angular misalignment. For the trap model studied here, which has similar dimensions to our fabricated trap, we find that the axial field can vary up to a few hundred V/m for linear misalignment $l$ up to 20 $\mu$m. This type of misalignment should be greatly suppressed in our trap, as the linear alignment is given by precisely machined alignment lobes. The linear alignment between wafers has been previously measured to be within 2 $\mu$m of the specifications \cite{simon2019}. 

\begin{figure}[h]
\centering
\includegraphics[scale=0.6]{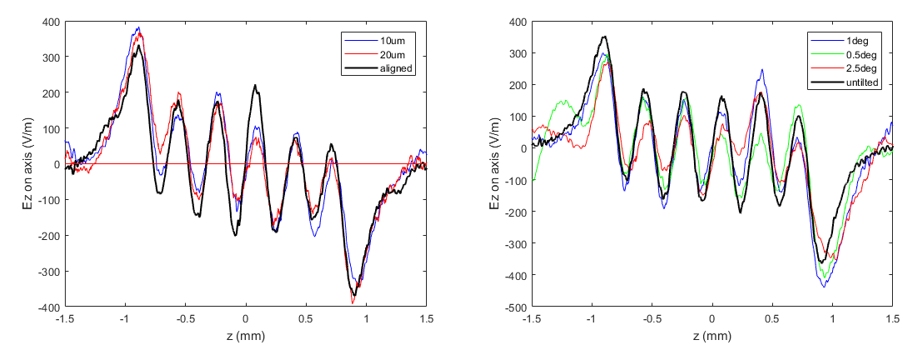}
\caption{Simulation of linear and angular misalignment on a 7-finger linear Paul trap. Left: linear misalignment up to 20 $\mu$m between the trap wafers. Right: angular misalignment between the top and bottom wafer of up to 2.5 degrees.}
\label{fig:misal_data}
\end{figure}

The angular misalignment has also been measured on test assemblies to be less than 0.05 degrees \cite{simon2019}, however due to the gluing process during assembly we may expect a high variation between different assemblies. Therefore we simulate the effect of angular misalignment of up to 2.5 degrees. Similarly to the case of linear misalignment, Figure $\ref{fig:misal_data}$, we find that fields on axis can vary up to a few hundreds of V/m for the larger 2.5 degrees misalignment. This study provides us with an estimate for potential residual axial fields along the trap axis, and can be incorporated in the study of axial micromotion. Our final study focuses on investigating the shaping of electrode fingers near the ion in order to minimise the residual axial field. We have seen from the simulation presented above that the axial field is minimal at the center of electrodes and at the center of gaps, however it can reach several hundreds of V/m fluctuations away from those locations. We are interested in assessing whether the axial field can be lowered at all locations by engineering electrode shapes which can cancel the residual axial effects. 
We begin our investigation by curving the electrodes edges as shown in Figure $\ref{fig:gaps_shaping}$ a). We find that this type of shaping raises the axial fields. Therefore we turn to smaller indentation features of different sizes. After exploring different shapes and sizes, we find that circular indentations with 5 $\mu$m in depth and 60 $\mu$m in width do minimise the residual axial field. This study shows that the features required to cancel out the existing fields are very small, and in the order of a few micrometers in size. This is a similar characteristic dimension to potential machining and fabrication defects.

\end{document}